# Dataset: Data Collection and Analytics from Soil Sensors and Weather Stations at Production Farms

Charilaos Mousoulis, Pengcheng Wang, Nguyen Luu Do, Jose F Waimin, Nithin Raghunathan,
Rahim Rahimi, Ali Shakouri, and Saurabh Bagchi

Purdue University

## ABSTRACT

Weather and soil conditions are particularly important when it comes to farming activities. Study of these factors and their role in nutrient and nitrate absorption rates can lead to useful insights with benefits for both the crop yield and the protection of the environment through the more controlled use of fertilizers and chemicals. There is a paucity of public data from rural, agricultural sensor networks. This is partly due to the unique challenges faced during the deployment and maintenance of IoT networks in rural agricultural areas. As part of a 5-year project called WHIN we have been deploying and collecting sensor data from production and experimental agricultural farms in and around Purdue University in Indiana. Here we release a dataset comprising soil sensor data from a representative sample of 3 nodes across 3 production farms, each for 5 months. We correlate this data with the weather data and draw some insights about the absorption of rain in the soil. We provide the dataset at: https://purduewhin.ecn.purdue.edu/dataset2021.

## 1 INTRODUCTION

The popularity of Internet of Things (IoT) [2, 6] sensors has pushed the market towards the commercialization of a large variety of low-power wireless sensors and cloud services for their integration, visualization, and analytics [10, 12]. Adoption of these IoT ecosystems has rapidly increased in industrial and household settings. Moreover, manufacturers begin to equip their production lines with sensors to allow for predictive maintenance and cost analysis, with the goal of reduction of unexpected downtimes and more efficient resource management [4]. However, the integration of IoT sensors in agriculture has been slow. This can be attributed to practical issues: i) limited Internet and broadband connectivity in large rural farms, ii) node communication issues due to ground morphology, large distances, and varying conditions throughout the season (e.g., growing crops), iii) sparsity of powered locations for gateways and IoT hardware, iv) access difficulty for on-site troubleshooting and maintenance, and v) lack of trained manpower that is often still needed for updating and maintenance of IoT systems.

Furthermore, there can be unique distinctive properties between adjacent fields or fields within the same IoT sensor network, such as: i) soil types and cover crops, ii) planted crops, iii) treatment strategies including applications of fertilizers, pesticides and insecticides, and more [3, 5, 7]. Therefore, data and insights collected from one type of field might not be applicable to neighboring fields.

The value of the insight from these types of data is very important to the farmer as well as for environmental sustainability. In the former case, information of soil moisture content over time and after significant weather events (e.g., extended rainfall or high temperatures) can provide insight on the absorption of nutrients and chemicals at different depths [11]. Since the cost of farming chemicals, such as fertilizers, can be relatively high, an indication of absorption and retention can lead to profitable financial decisions on the purchase of the amount that is actually needed in each case. On the other hand, avoidance of over-application is important for environmental sustainability, since the excess chemicals often end up being washed into our water bodies.

Our provided dataset aims to fill the gap of real-world measurements that can be used for precision agriculture. The dataset comes from IoT nodes with temperature, humidity, and soil sensors (measuring temperature, conductivity, and volumetric water content). In addition, the dataset contains data from adjacent weather stations providing various measurements, including rain events. We have provided measurements collected by three nodes and two weather stations over a duration of five months. Each node is located at a different property (field), which have cultivation of soy or corn. We have publicly released the dataset at https://purduewhin.ecn.purdue.edu/dataset2021.

## 2 OVERVIEW

In this section, we describe the hardware setup for our nodes as well as the deployment information.

### 2.1 Hardware Setup

Our teams have developed an embedded system for the acquisition and transmission of both commercial and custom sensor measurements. The system is based on the nRF52832 SoC from Nordic Semiconductor with the addition of a LoRa module from Semtech. The node, shown in Figure 2, is an improvement over the version published earlier [8] and has external connections for 8 sensors. Currently, we have allocated half of these connections for analog sensors with a 24bit ADC module and the other half for digital sensors through a serial interface. For the described dataset we present the measurements from deployed nodes with the Teros-12 commercial soil sensors from the Meter Group [9].

The communication of the nodes to the gateways is through a custom LoRa-based protocol that we design to increase the reliability and the reach of the networks. The transmission is at 915 MHz with a spreading factor equal to 9. The purpose of this custom communication protocol is to allow for mesh operation and therefore increase significantly the range of the deployed nodes. Our gateways are Raspberry Pi equipped with a LoRa transceiver (LoStik, Ronoth) which upload the measurements to our databases through a wired network. A large amount of the received data is available to the community through our open portal. The portal provides visualizations and downloads of current and historical data from the nodes that are deployed in the community. A geo-tagged list of nodes along with their metadata (e.g., timestamp of last received transmission and battery level) is provided [1], from which the visitor can click on a specific node and access its measurements both graphically and through a downloadable csv format.

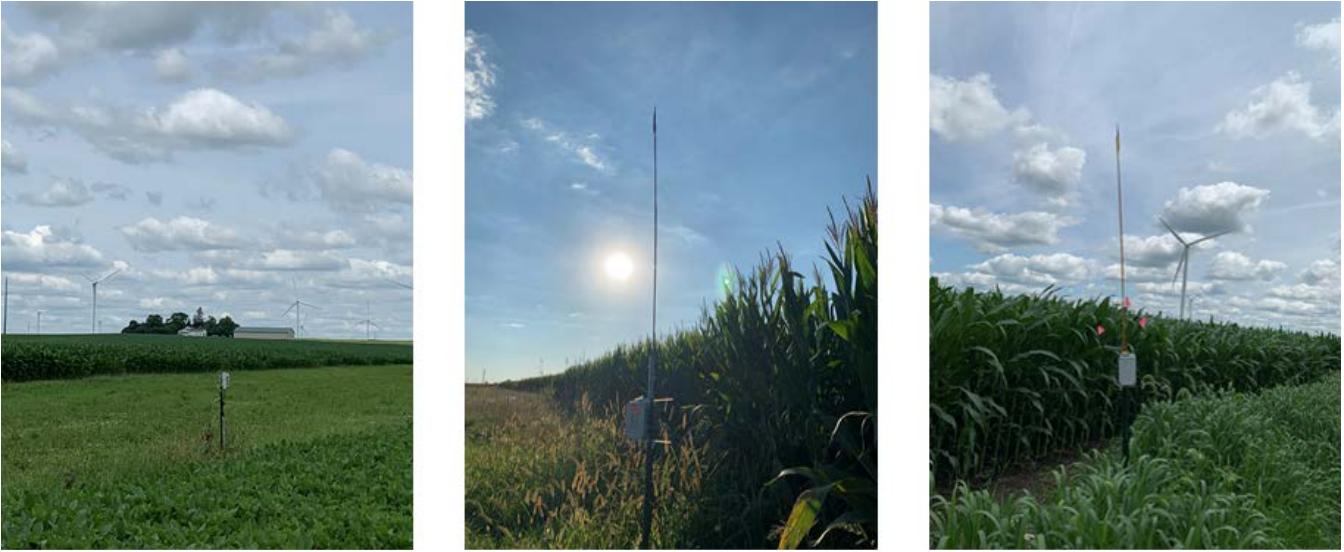

*Figure 1: Deployments of the LoRa-based soil-sensing nodes at different fields. Left: Soybean field; Center and right: Corn fields. The nodes at the corn fields have their antennas mounted at higher points due to the height of the grown crop. In addition to the temperature, volumetric water content, and conductivity of the soil, the nodes are transmitting the ambient temperature and humidity.*

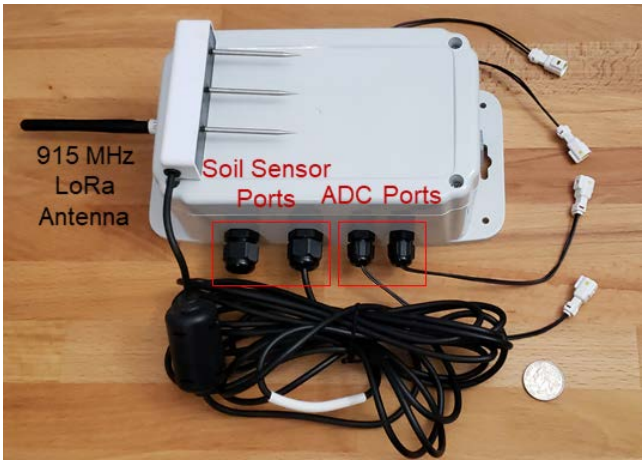

*Figure 2: Photo of the node with one soil sensor connected. The 915 MHz LoRa antenna as well as the soil sensor and ADC ports are visible in this view. There are two more pairs of ports on the other side.*

## 2.2 Deployment Information

The goal of our deployments and the broader WHIN project is to make the 10-county region in the heartland of Indiana a hub for precision agriculture, enabled by novel IoT, AI, and our low-cost sensor fabrication technologies. We therefore work with farmers in the region for the deployment of the IoT sensor networks that communicate to our central framework for visualization and analytics. For each node we use two soil sensors that are located at 6- and 12-inch depths.

Each soil sensor provides the volumetric water content (VWC; $m^3/m^3$), temperature ($°F$), and conductivity ($dS/m$). To allow for the LoRa communication, the nodes are mounted high on metallic posts. The gateways are positioned within range and at locations where there is available power. Typically, this is at a shed, office, or the farmer's residence. There is a need for internet connectivity at those points, but since the transmission rate is low (i.e., typically every few minutes) and the packets are small, broadband connection is not required.

In addition to the soil information collected by the connected sensors, our nodes transmit temperature and humidity (%RH) information as collected by an on-board sensor. This allows for the comparison of the soil temperatures to the ambient conditions which are registered inside the enclosure of our electronics. Even though the enclosure features a vent membrane, it has been designed to withstand complete immersions in water for a short period of time (IP68 rating).

## 3 COLLECTED DATASET

The dataset contains measurements collected by three nodes and two weather stations over a duration of five months. Each node is located at a different property (field) as shown in Fig. 1. Two of the fields have corn cultivation and the third has soy cultivation. The two weather stations are adjacent to the locations of the deployments. We have made the dataset public at https://purduewhin.ecn.purdue.edu/dataset2021. This dataset is categorized into folders with the following structure and name:

(1) Soil-sensing LoRa nodes - Farmers 1 and 2. Farmer 1 grows soy, whereas Farmer 2 grows corn. The nodes are at a distance of 0.85

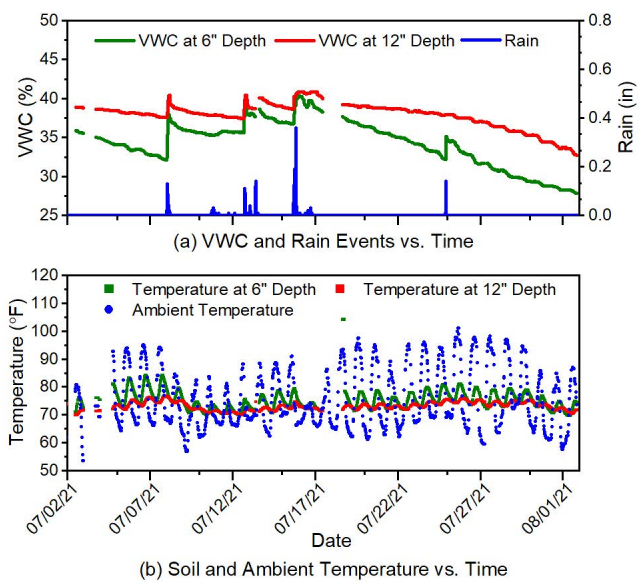

*(a) VWC and Rain Events vs. Time*

*(b) Soil and Ambient Temperature vs. Time*

*Figure 3: Filtered data from LoRa node 1 and weather station 1: a) Volumetric water content values for the two depths and rain events for a duration of approximately 1 month, b) temperatures at the different depths as registered from the node for the same duration.*

miles (1.36 km) between them but are in different properties. The farmer data timestamps are in Eastern Time (ET) timezone with 30 min interval. Measurement data include i) battery voltage (*V*), ii) ambient temperature, iii) ambient humidity, and iv) temperature, volumetric water content, and conductivity for sensor 1 at 12-inch depth and sensor 2 at 6-inch depth.

(2) Soil-sensing LoRa node - Farmer 3. Farmer 3 grows corn and is at a different location than Farmers 1 and 2. The provided data are as described in the previous case.

(3) Weather Station 1 and 2. Weather station 1 is located at a distance of approximately 1 mile (1.6 km) from the nodes of Farmers 1 and 2. Weather station 2 is at a distance of approximately 7.43 miles (11.96 km) from Farmer 3. The weather station data timestamps are in UTC timezone with 15 min interval. Measurement data include i) total rainfall measured during the interval and total rainfall measured during the last hour (in 1/100 of an inch increments), ii) average, low, and high surface temperature over the time interval (*°F*), iii) relative humidity and barometric pressure (*mmHg*), iv) average and gust wind speed (*mph*) as well as prevailing and gust wind direction (clockwise degrees), v) soil temperature in 2-, 5-, 10-, and 15-inch (*°F*), and vi) soil moisture at 2-, 5-, 10-, and 15-inch depths (centibar),

## 4   EXAMPLE ANALYSIS

One of the analyses that can be done is on the effect of rain events on the temperature and water content at different soil depths for a particular crop type. We show as a representative sample, the result of such analysis at one of our locations.

For this analysis we use part of data collected from the IoT node with the soil sensors at Farmer 1 (soy field) and plot the VWC, converted into percentages, on the Y-axis for the soil sensor at

6-inch depth and the one at 12-inch depth using the X-axis for time (dates). A percentile filter with moving average of 5 data points is applied to smooth the VWC data; the range of possible values is between 0 and 60%. Out-of-bounds values occur occasionally due to corrupted packets. We also plot the rainfall (over the 15-min intervals) on the secondary Y-axis (to the right) for the same duration. The rainfall is provided by the adjacent weather station 1. The graph is shown in Figure 3(a). Several interesting observations can be made by studying the trends of the plotted measurements, such as, a) the water content difference between the 2 depths can be up to 5%, b) even though rain events cause a noticeable variation at 6 inches, not all rainwater ends up at the 12-inch depth, and c) a higher rain event (e.g., the one occurred on 7/15) may not have as big an impact on the soil water content if the water content is above a certain level, in this case above 35%.

On Figure 3(b) the ambient and soil temperatures at the two different depths are shown for the same time duration. It can be noted that at the shallower depth the temperature variations are higher, as they are more affected by the environmental variations. However, at the greater depth the temperature has smaller variations throughout the day and also exhibits a small hysteresis. Furthermore, the relationship between the ranges and variations of the ambient and the soil temperatures can be deduced, providing valuable insights on the conditions that the crop roots are experiencing.

## 5   CONCLUSION

In this work, we present an agricultural soil and weather dataset collected with nodes and stations deployed in corn and soybean fields across two counties in north-central Indiana. Our datasets include measurements collected throughout the farming season, in soybean and corn fields, and show the variations and dependencies between the environmental and soil conditions. The datasets can be used for research as well as for decision making, such as to deduce the amount of fertilization to be applied and the ability of the soil to retain moisture at different levels.

## ACKNOWLEDGMENTS

We like to acknowledge the incorporation of WHIN https://www.whin.org weather station data into our dataset. WHIN's data is collected as part of a collaboration with farmers, manufacturers, and technology partners in the 10-county Wabash Heartland Living Lab ecosystem. The data is provided to promote and accelerate research and education in STEM-related fields and IoT. The WHIN project is supported by a 5 year, $40M grant from the Lilly Endowment Inc. (2017-22). We would also like to thank Leslie Fisher for her help with farmer engagement and during the actual field deployments.